# A Novel Handover Mechanism for Visible Light Communication Network

**M.A.N. Perera, N.G.K.R. Wijewardhana, A.A.D.T Nissanka, S.A.H.A. Suraweera and G.M.R.I. Godaliyadda**

**Abstract:** Visible light communication (VLC) is an emerging technology and considered as an alternative to overcome some of the disadvantages of radio frequency communication technology in an indoor environment. However, the line of sight nature of the technology limits the user mobility and create new challenges to provides seamless network coverage under user mobility scenarios. In VLC multi access points and multi cell based network, co channel interference (CCI) between neighbor cells limits the overall performance. Therefore, by following the already existing VLC handover systems, this statistical parameter based novel handover method is designed. This paper proposes a handover mechanism for indoor VLC systems by introducing cell ID bits and statistical kurtosis values of the cell ID waveforms as metric for the handover initiation. As advantage of the method, effects from the CCI can be eliminate when measuring signal strengths in the signal overlapping cell boundary area. Also the handover criteria are adoptive to the different ambient lighting conditions compared to existing pre-configured light intensity threshold based handover systems. With the use of bit error rate (BER), the experiment results showed that Kurtosis value of the cell ID waveforms can be used as metric to initiate network handover in indoor VLC systems

## 1. Introduction

Visible light Communication (VLC) has become a popular research area during recent years, due to the lack of bandwidth in the frequency spectrum for radio frequency (RF) communication [1,2]. The bandwidth frequency spectrum of visible light is larger than that of RF bandwidth. RF bandwidth ranges from 3 kHz to 300 GHz and VLC spectrum offers a bandwidth range from 400 THz to 800 THz. There are several other advantages in VLC compared to RF based mobile communication systems. Unlike RF mediums, visible light cannot penetrate walls in indoor environments. This characteristic provides an inherent network security for indoor environments. The use of visible light as a carrier of data reduces the VLC systems from electromagnetic interference and associated health concerns. Therefore, VLC systems can be implemented on places where RF communication systems have disadvantages. Such as hospitals, factories, under water, inside airplanes etc. Also VLC facilitates the reuse of existing light configuration for communication. [3] Since the characteristics of reusability and abundant spectrum compared to RF, VLC systems can be built with lesser efforts and at a lower cost.

In order to use VLC in mobile network systems, it is necessary that this system can provide uninterrupted high speed connectivity in presence of user mobility scenarios. To use VLC systems as an alternative for RF communication systems, it should be able to initiate handovers between access points to provide seamless connectivity for the mobile users [4]. Usually in RF communication systems handover is initiated on the basis of

*Mr. M.A.N Perera,* *B.Sc.Eng.(undergraduate) Department of Electrical and Electronics Engineering, Faculty of Engineering, University of Peradeniya.*

*Mr. N.G.K.R. Wijewardhana,* *B.Sc.Eng.(undergraduate) Department of Electrical and Electronics Engineering, Faculty of Engineering, University of Peradeniya.*

*Ms. A.A.D.T Nissanka,* *B.Sc.Eng.(undergraduate)Department of Electrical and Electronics Engineering, Faculty of Engineering, University of Peradeniya.*

*Dr. S.A.H.A. Suraweera,* *B.Sc.Eng.(Peradeniya), Ph.D. (Monash University), Senior Lecturer, Department of Electrical and Electronics Engineering, Faculty of Engineering, University of Peradeniya.*

*Eng. (Dr.) G. M. R. I. Godaliyadda,* *B.Sc.Eng.(Peradeniya), Ph.D. (National university of Singapore), Senior Lecturer, Department of Electrical and Electronics Engineering, Faculty of Engineering, University of Peradeniya.*

physical parameter measurements, such as received signal strength. To define optimal criteria for handover, there are several factors taken into consideration. such as energy consumption, positioning, end user quality of service etc.

This paper proposes a handover method for indoor multi cell based VLC networks. Cell ID bits and statistical parameters are introduced to eliminate the CCI. In indoor VLC systems, Intensity level can vary quickly due to reasons such as receiving height, receiving orientation, background light intensity etc. As an advantage of using statistical parameters, impact from above mentioned incidents can be minimized compared to other intensity threshold based handover methods. According to the purposed methodology, received visible light cell ID wave form is analyzed statistically. Handover initiation is decided when statistical parameters reached to their relative maximum value. Since there isn't any comparison between pre-configured constant threshold values, the system can be designed to adopt in different uniform lighting conditions.

The rest of the paper is organized as follows. Section 2 describes the existing methodologies for VLC networks and their drawbacks. The system model is explained in section 3. Section 4 gives the experiment details and observed results. Finally, conclusions are given in section 5.

## 2. Literature Review

Presently there are several technologies designed for achieving seamless connectivity for mobile users [5,6,7]. One of the most widely recognized standardized approaches to VLC is the exploitation of IEEE 802.15.7 standard for visible light data communication. IEEE 802.15.7 [8] supports high-data-rate visible light communication up to 96 Mb/s by fast modulation of optical light sources which may be dimmed during their operation. IEEE 802.15.7 provides dimming adaptable mechanisms for flicker-free high-data-rate visible light communication.

In [9] vertical handover initiating algorithm is proposed for VLC – LTE Hybrid communication system. since this method designed for hybrid communication system, it cannot apply on handovers in VLC only systems. In [10] authors propose two handover mechanisms for overlapped uniform VLC cells and non – overlapped spotlight VLC cells. In overlapped uniform VLC cell, the user performs intensity test and detects the presence of overlapped area [11]. It initiates the handover when intensity reach the threshold level. In spotlight VLC cell method, there is buffer memory allocated to save data on mobile user to avoid connectivity interruptions in dead zones. In [12], there is an optical at to cell network proposed which can reduce the co channel interference by using joint transmission method from multiple access points. However, as the authors mentioned the downside of joint transmission systems is that they need extra signaling overhead. In a cell based network, using same frequency in adjacent cells causes co-channel interference [13]. This significantly reduce the network performance to a mobile user in overlap region. Therefore, interference coordination techniques are used to eliminate CCI. In [14] CCI is eliminated by splitting cells into clusters. Inside of a cluster, frequency resources are used in orthogonal manner.

## 3. System Model

The system model consists of multiple transmitters and multiple receivers. The block diagram of a single data link is shown in Fig. 1.

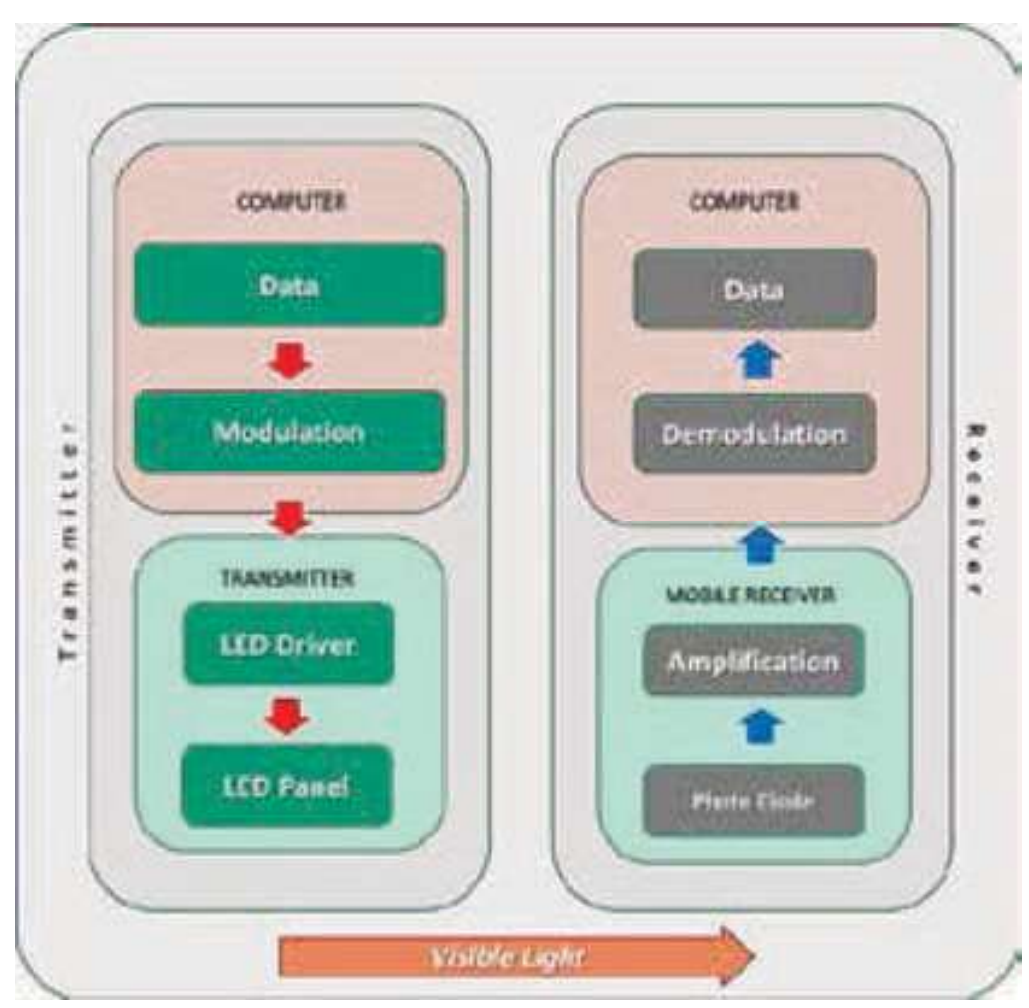

**Figure 1 – Block diagram of system model.**

### 3.1 Transmitter

The transmitters were designed to provide good indoor illumination in addition to data transmission. Data communication by using higher illumination intensities can minimize the distortions to the transmitted data in the transmitting medium.

The Prototype transmitter unit consists of LED driver and power LED panel. Since power LEDs have the high speed switching capability and higher illumination intensity compared to the traditional LEDs, 5x 7w Chips on Board type (COB) cool white power LEDs were used in order to achieve large distance data transmission. COB power LED contain more light sources in the same area than standard LEDs could occupy. This will result a greatly increased illumination output per square meter.

In order to handle required current for the LED panel, TTL compatible metal–oxide–semiconductor field-effect transistors (MOSFET) were used. Transistor switching circuit was used for the LED driver. The MOSFETs were operated by switching signal from the Arduino output. The switching signal carries the data which was needed to be transmitted. The block diagram and the prototype transmitter are shown in Figure 2 & Figure 3.

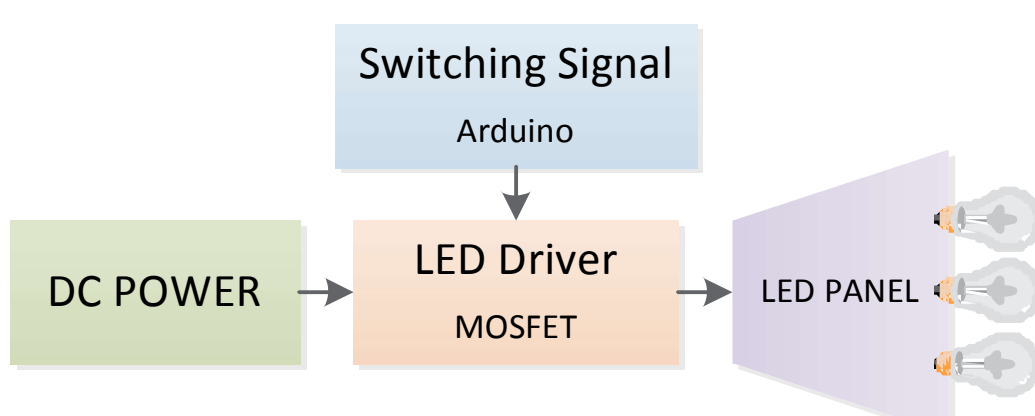

**Figure 2 – Block diagram of prototype receiver.**

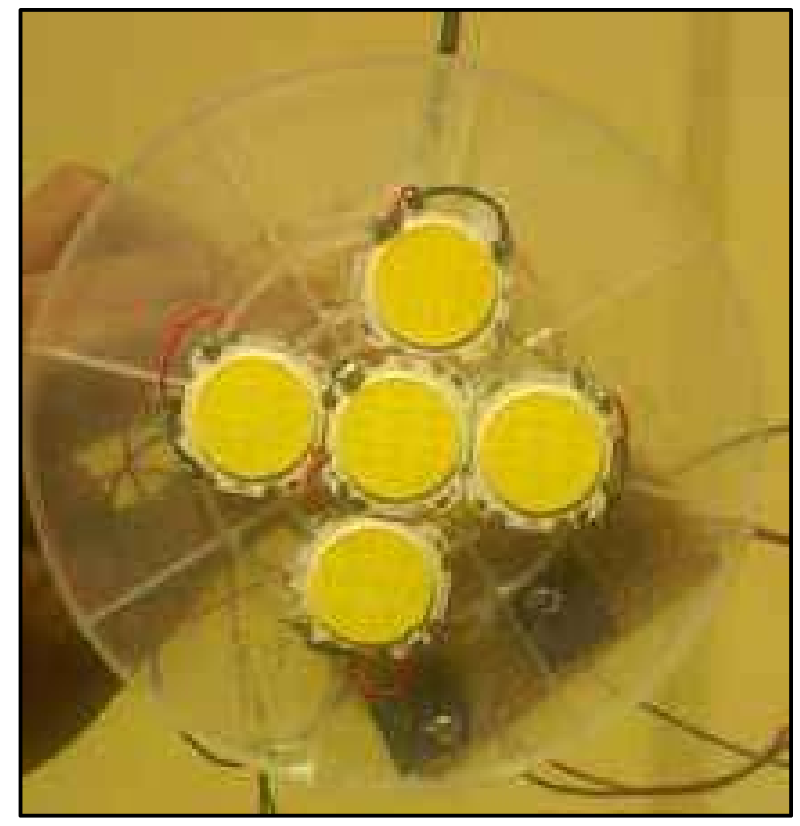

**Figure 3 – Prototype transmitter with multiple power LEDs.**

### 3.2 Receiver

Typically, in indoor environments, there is a highly possibility of distorting and interfering the incoming signal wave due to the DC components from ambient light and rectified sine wave emitted from filament, fluorescent lamps. [16] In order to minimize the impact from background disturbances, two photodiodes (PD) with noise cancellation method was used in the receiver unit. Two transimpedance amplifiers were used to convert PD's output current in to voltage signal. DC shift corresponding to ambient noise and light sources were reduced by using differential amplifier in the receiver. Also a reflector fixed on signal receiving PD to improve the signal receiving vertical distance. The reflector directs the signal receiving PD to transmitter direction. Therefore, it limits the signal receiving PD scope and minimize the exposure of the ambient noise. To further minimize the DC shifts of the incoming waveforms, a high pass filter was used in the receiver. The prototype receiver is shown in Figure 4.

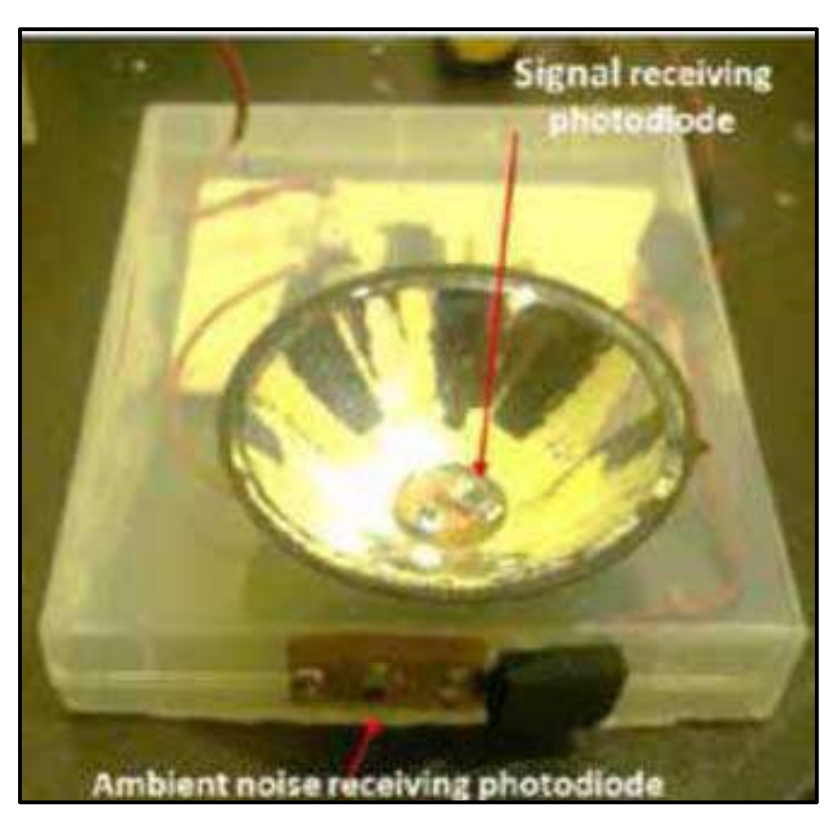

**Figure 4 – Prototype receiver.**

Then, received data is transmitted to receiver's ADC unit to sample the waveforms at higher sample rate. Block diagram of the prototype receiver is shown in Figure 5.

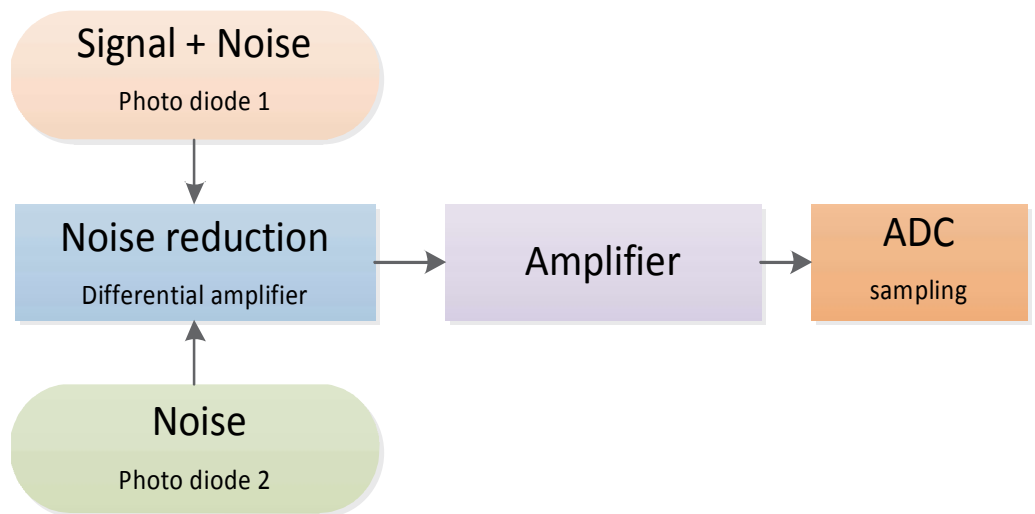

**Figure 5 – Block diagram of prototype receiver.**

Two ARDUINO UNO boards were used to the communicate with the prototype VLC system.

### 3.3 Handover Initiation

In VLC system model can be described as illustrated in Fig. 6. To initiate handover, system follow few steps.

- A data structure is defined to transmit data from transmitter to receiver. First two bit of the data structure is used for cell ID bit. each transmitter adds its cell id bits before data transmission [17]. The cell ID bits are transmitted by OOK (On Off Keying) modulation. Fig. 7 illustrates the received data frame structure.

- Then from the receiver, received wave form shape is analyzed. Mobile user separates the cell id segment and data segment of the data structure. Then it calculates the kurtosis value of the cell ID waveform to estimate position. As shown in figure, different shapes of cell id waveforms received for different user positions. Fig. 7 illustrates the received data frame structure.

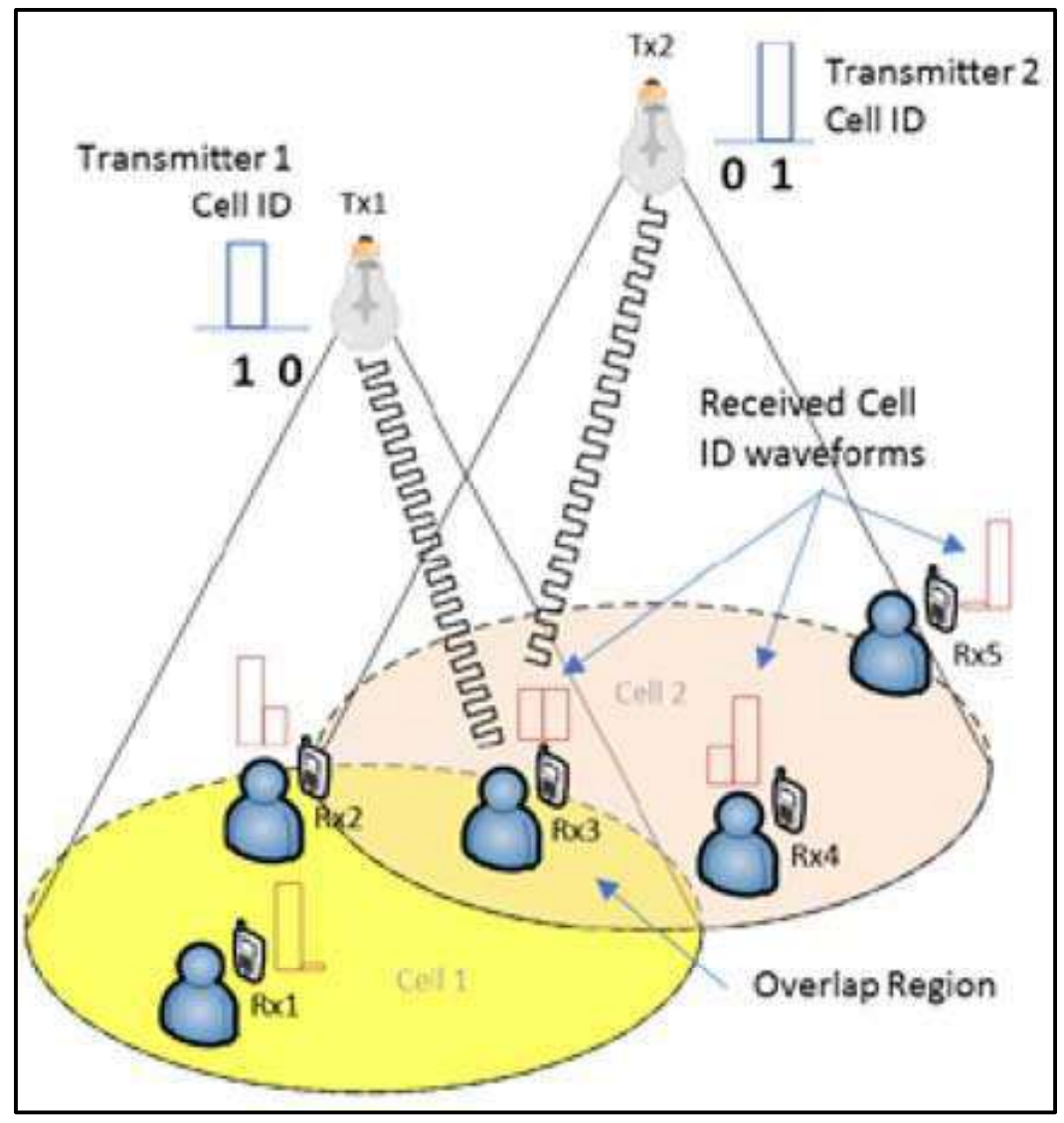

**Figure 6 – VLC system with cell ID waveforms.**

- Handover initiation decision is based on statistical characteristics of the ID wave form. According to the Fig. 6, Rx3 receives superimposed signals which have equal signal strengths on two cell bits while, Rx5 and Rx1 receive only one strong signal on cell bits. The waveform shape depends on the received signal strength of the transmitted cell ID. Received signal is shown in Figure 7.

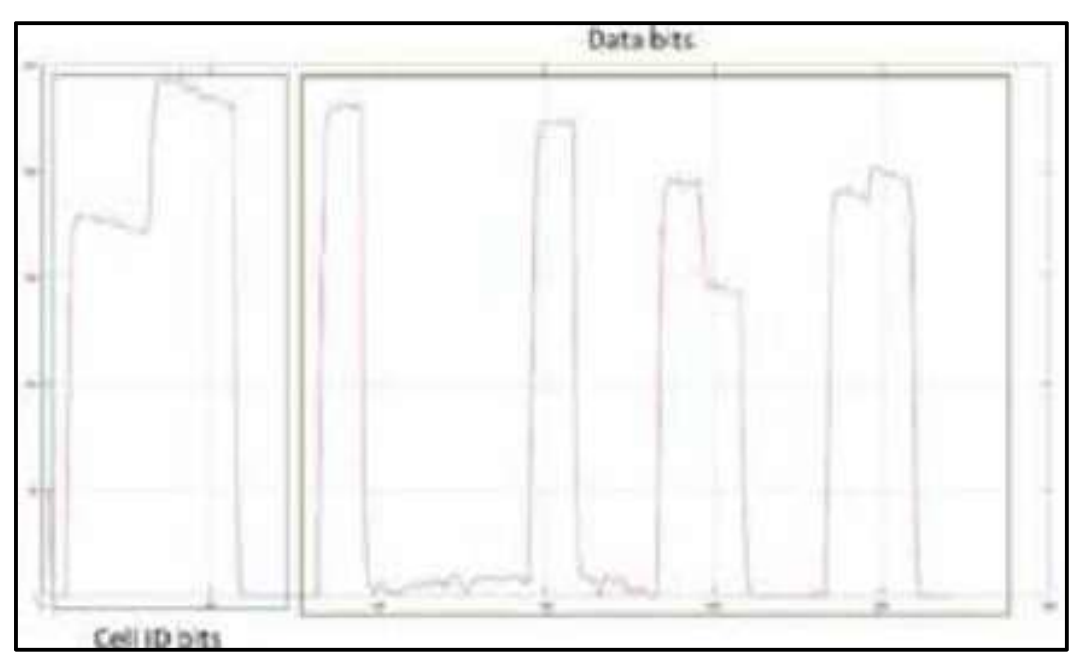

**Figure 7 – Received data structure.**

### 3.4 Mathematical Background

The shape of cell ID waveform varies with the received cell ID signal level strengths of adjacent transmitters**.** The variations of this type waveforms can be analyzed by calculating the kurtosis value of the waveform. User position for the handover initiation can be estimated based on the kurtosis values. For the calculation of the kurtosis, conditional probability density function (PDF) of the waveform was constructed using sampled data of the received waveform. Mathematically, the sampled values constructed from the received waveform from the PD, $y(t)$ can be written as

$$y[n] = y(nT_s) \quad n = 1,2,3 \ldots \qquad \ldots(1)$$

where $T_s$ is the sampling time and $y[n]$ is the sampled value.

The sample values are mapped to a histogram to construct the PDF of distribution.

Kurtosis is a measure of whether the data are heavy-tailed or light-tailed with respect to a normal distribution. The kurtosis decreases as the tails become lighter and it increases as the tails become heavier. Kurtosis for a discrete set of data is

$$\text{Kurtosis} = \frac{\sum_{i=1}^{N}\left(Y_i - \bar{Y}\right)^4 / N}{s^4} \qquad \ldots(2)$$

Where $Y$ is sampled data point, $\bar{Y}$ is mean of sampled data, $S$ is standard deviation and N is the sample size.
If the kurtosis is close to 0, then a normal distribution is often assumed. These are called *mesocratic* distributions. If the kurtosis is less than zero, then the distribution is light tails and is called a *platykurtic* distribution. If the kurtosis is greater than zero, then the distribution has heavier tails and is called a *leptokurtic* distribution [13].

Kurtosis is a component of both tailedness and peakedness of a distribution. It is basically because kurtosis represents a movement of mass that does not affect the variance. In the case of positive kurtosis, where heavier tails are accompanied, can be modeled to the waveform received for the cell identification on handover region. If the distribution is more concentrated towards the tails, then the variance will also be larger. To leave the effect of variance unchanged, standardized kurtosis is taken with respect to variance. Hence a high value for kurtosis is obtained at the handover position. It is also recognized that although tailedness and peakedness are often both components of kurtosis, kurtosis can also reflect the effect of primarily one of these components, such as light tails. Thus, for symmetric distributions, positive kurtosis indicates an excess in either the tails, or peakness which is used to identify the receiver location.

In this handover method, combination of kurtosis and bit error rate is used to initiate handover, hence a more reliable outcome is achieved. Bit error rate is calculated from the equation,

$$BER = \frac{No.of\ bits\ in\ error}{No.of\ total\ bits\ transmitted} \qquad ..(3)$$

## 4. Experiment & Results

### 4.1 System Setup

The experiment was carried out in an area of dimensions 1.6 m x1.6 m. The experiment was carried for two transmitters and one mobile observer. Observed user moving path across boundary of the neighboring cells is shown in Fig. 8. Two transmitters are named as T1 and T2 in the figure. BER was calculated by using 8 x 8 grid with resolution of 0.2m x 0.2m placed over the area. Transmitters were fixed 2m above from receiver plane. Transmitters were the only light sources at the experiment premises.

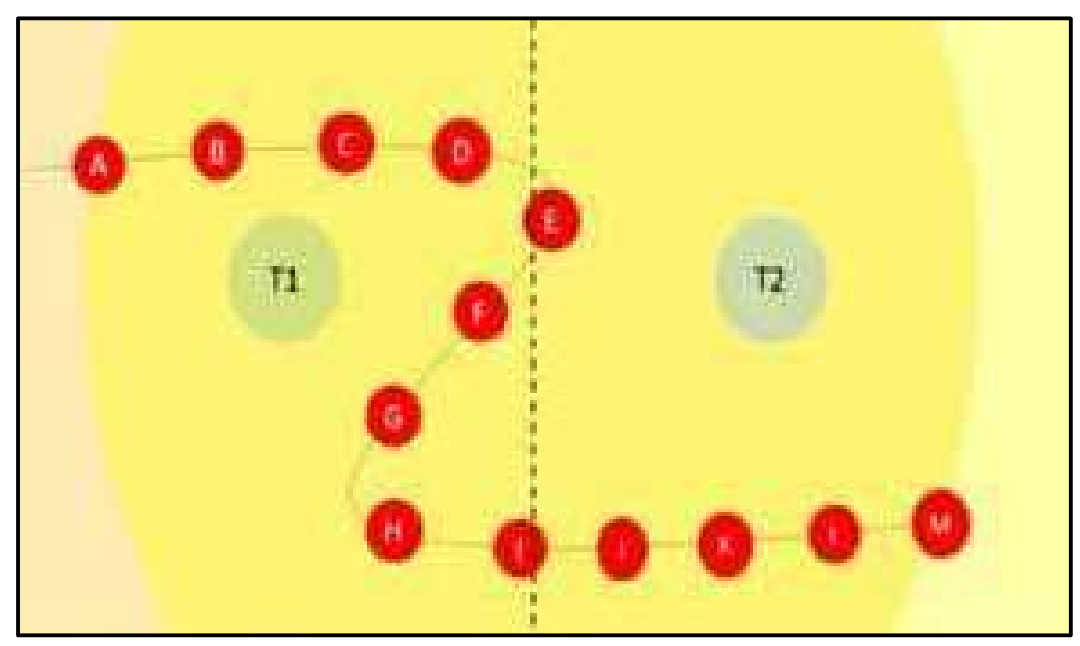

**Figure 8 – Path of the mobile user**

### 4.2 Results

To observe the receiving cell ID waveform patterns, 0 1 and 1 0 bit sequences transmitted as cell ID from T1 and T2 respectively. Fig. 9 shows the received Cell id waveforms when moving along A to M. Random walking pattern across the two transmitters was taken to consideration when creating the observing path. From position E to H user change the moving direction and from H onwards it continues to go across the two transmitters. Calculated kurtosis values variation for the received cell ID waveforms is given in Fig. 10. From the received waveforms shown in Fig. 9, it can be observed that waveforms corresponding to position (E) and (I) have the highest kurtosis values compared to other positions.

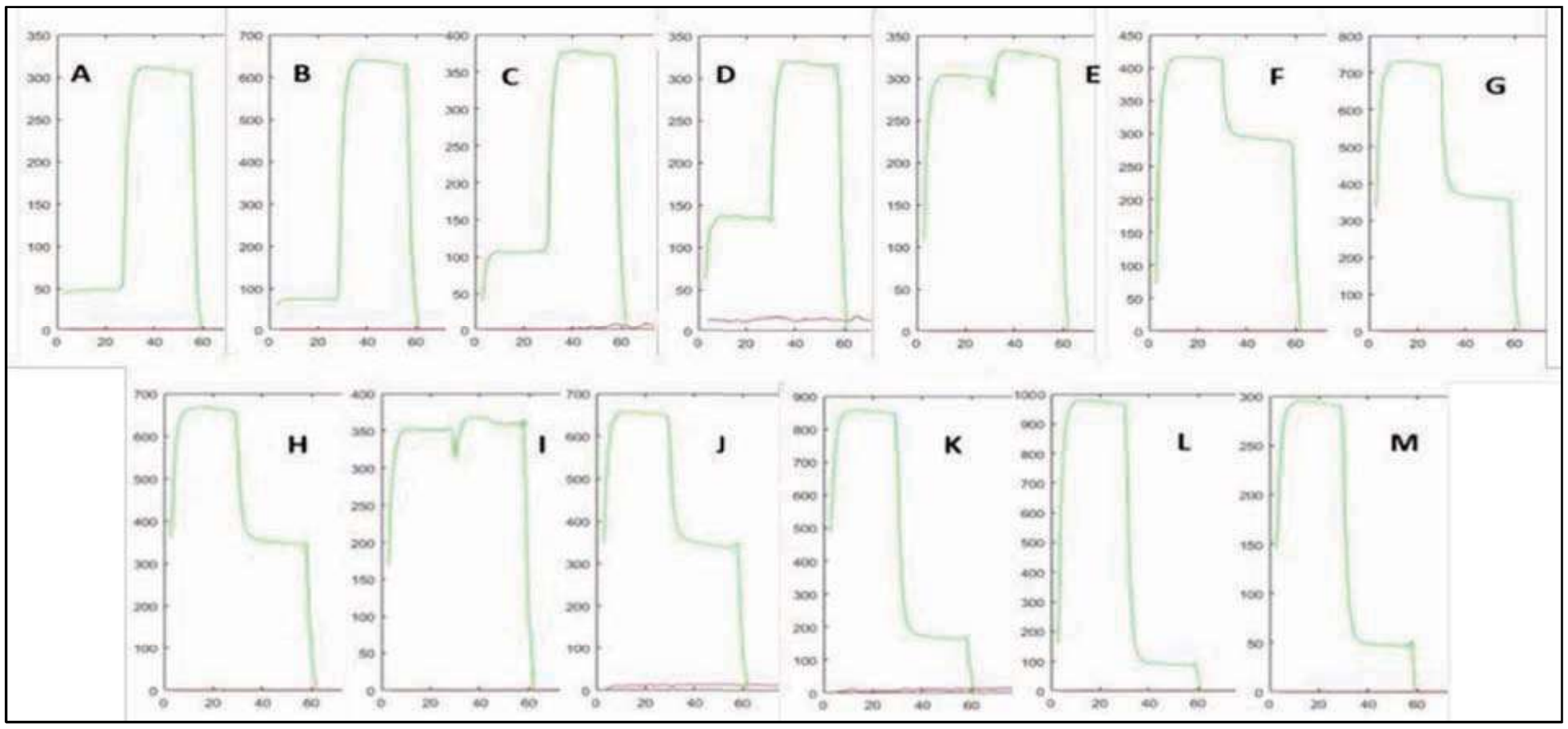

**Figure 9 – Received cell ID wave form for different positions.**

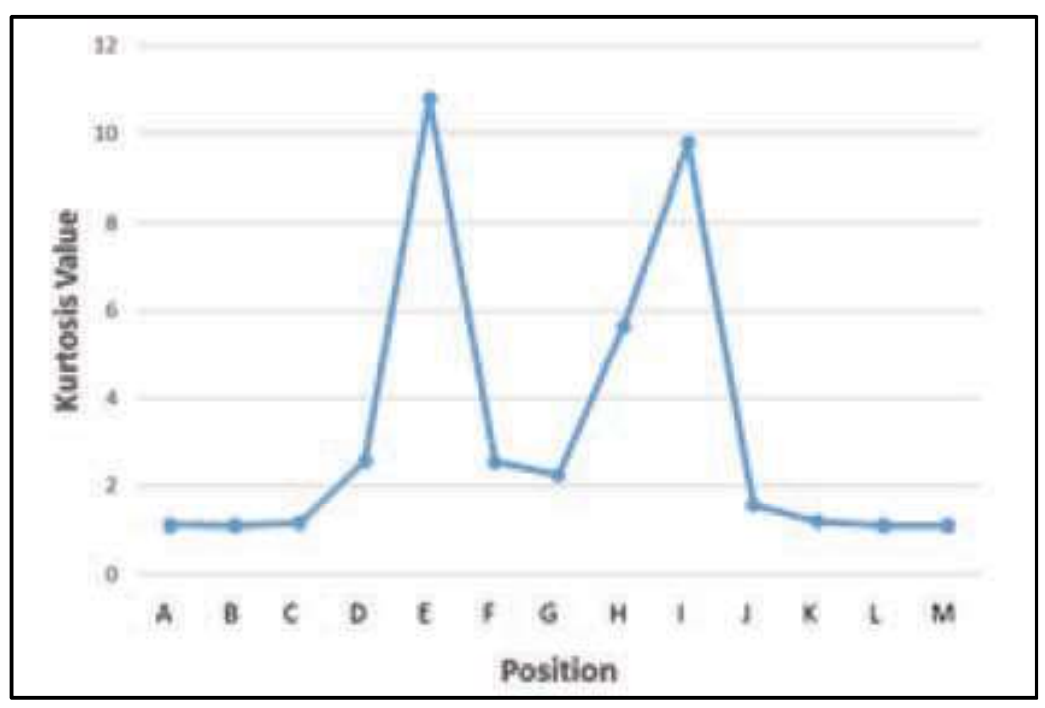

**Figure 10 – Kurtosis values for different positions.**

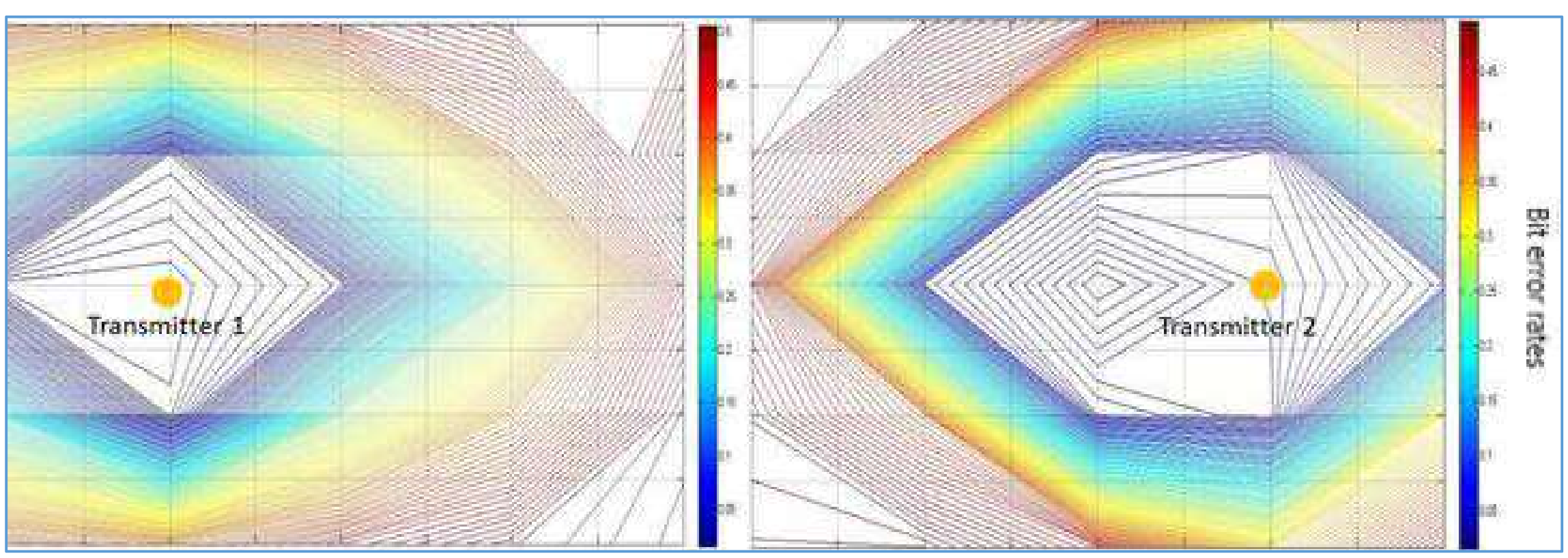

**Figure 11 – BER for TX1 & TX2 in different positions of the grid.**

Measured BER on different positions of the 8x8 grid is shown in Fig. 11. Each coloured contour lines represents a different BER value observed in the area.

### 4.3 Handover Based on Results

Fig.12 shows the BER variation with kurtosis along the A-M path. Position (E) and (I) have the minimum BER in both TR1 and TR2. Positions directly under the transmitters have low kurtosis values compared to the other positions. Such as position A, B, C, G, J, K, L, M. Therefore, it can be clearly observed that positions which have higher CCI results high Kurtosis values. According to the results position (I) and (E) which have equal signal strengths from the TR1 & TR2, can be used as a handover initiation positions. However, since user change its moving direction, handover initiation is not required in position E.

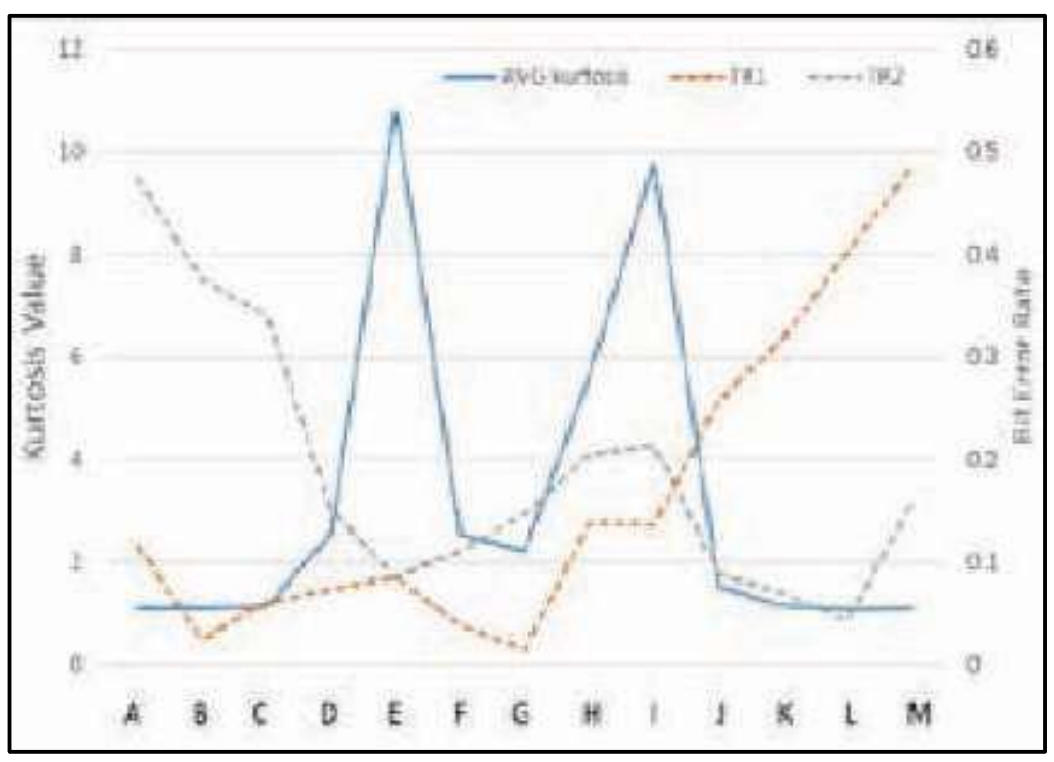

**Figure 12 – Kurtosis values and BER variations for different positions**

Therefore, experiment results shown in Fig. 12 concludes that higher kurtosis values of the cell ID bits directly related to the minimum BER positions of the multiple access points. Results of the prototype experiment demonstrate that kurtosis parameter can be used as metric for handover position together with minimum bit error rate.

## 5. Conclusions

This paper presents a novel handover mechanism for VLC network in order to mitigate CCI in neighbor cells and detect the accurate signal strengths of the neighbor transmitters in order to initiate handovers and achieve seamless coverage for the mobile users. To perform that, cell ID bits were introduced and received waveforms were analyzed statistically. The kurtosis values and bit error rates from two transmitters were considered in the experiment and the results showed that the higher kurtosis values can be obtained when multiple transmitters have equal visible light signal intensity. Also low BERs were observed at the high kurtosis positions. Therefore, it can be showed that kurtosis values of the cell ID can be used as handover decision making metric for the indoor VLC networks. The system can be further improved in combination with other devices such as CCTV video cameras in order to track the walking patterns of the users and optimize the handover decision criteria.